# Data Science in Economics

**Saeed Nosratabadi** [1], **Amir Mosavi** [1,2]*, **Puhong Duan** [3] **and Pedram Ghamisi** [4]

1 School of the Built Environment, Oxford Brookes University, Oxford, OX3 0BP, UK.
2 Faculty of Civil Engineering, Technische Universität Dresden, 01069 Dresden, Germany.
3 College of Electrical and Information Engineering, Hunan University, Changsha 410082, China.
4 Exploration Devision, Helmholtz Institute Freiberg for Resource Technology, Helmholtz-Zentrum Dresden-Rossendorf, Dresden, Germany.
* Correspondence: a.mosavi@brooks.ac.uk;

**Abstract:** This paper provides the state of the art of data science in economics. Through a novel taxonomy of applications and methods advances in data science are investigated. The data science advances are investigated in three individual classes of deep learning models, ensemble models, and hybrid models. Application domains include stock market, marketing, E-commerce, corporate banking, and cryptocurrency. Prisma method, a systematic literature review methodology is used to ensure the quality of the survey. The findings revealed that the trends are on advancement of hybrid models as more than 51% of the reviewed articles applied hybrid model. On the other hand, it is found that based on the RMSE accuracy metric, hybrid models had higher prediction accuracy than other algorithms. While it is expected the trends go toward the advancements of deep learning models.

Acronyms

| Acronym | Explanation |
|---|---|
| (2D)2PCA | 2-Directional 2-Dimensional Principal Component Analysis |
| AE | Auto-Encoder |
| ANN | Artificial Neural Network |
| AO | Adam Optimizer |
| ARIMA | Autoregressive Integrated Moving Average Model |
| BRT | Boosted Regression Tree |
| CNN | Convolutional Neural Network |
| DCNN | Deep Convolutional Neural Network |
| DL | Deep Learning |
| DNN | Deep Neural Network |
| DNNC | Deep Neural Network Classifier |
| DRL | Deep Reinforcement Learning |
| EL | Ensemble Learning |
| ELM | Extreme Learning Machine |
| EWT | Empirical Wavelet Transform |
| GARCH | Generalized Autoregressive Conditional Heteroskedasticity |
| GRDH | Group Method of Data Handling |
| GRNN | Generalized Regression Neural Networks |
| GRU | Gated Recurrent Unit |
| KNN | K-Nearest Neighbors |
| LOG | Logistic Regression Classifier |
| LSDL | Large-Scale Deep Learning |
| LSTM | Long Short-Term Memory |
| LWDNN | List-Wise Deep Neural Network |
| MACN | Multi-Agent Collaborated Network |
| MB-LSTM | Multivariate Bidirectional LSTM |
| MDNN | Multilayer Deep Neural Network |
| MFNN | Multi-Filters Neural Network |
| MLP | Multiple Layer Perceptron |
| MLP | Multi-Layer Perceptron |
| NNRE | Neural Network Regression Ensemble |
| O-LSRM | Optimal Long Short-Term Memory |
| PCA | Principal Component Analysis |
| pSVM | Proportion Support Vector Machines |
| RBFNN | Radial Basis Function Neural Network |
| RBM | Restricted Boltzmann Machine |
| REP | Reduced Error Pruning |
| RF | Random Forest |
| RFR | Random Forest Regression |
| RNN | Recurrent Neural Network |
| SAE | Stacked Autoencoders |
| SLR | Stepwise Linear Regressions |
| SN-CFM | Similarity, Neighborhood-Based Collaborative Filtering Model |
| STI | Stock Technical Indicators |
| SVM | Support Vector Machine |
| SVR | Support Vector Regression |
| SVRE | Support Vector Regression Ensemble, |
| TDFA | Time-Driven Feature-Aware |
| TS-GRU | Two-Stream GRU |
| WA | Wavelet Analysis |
| WT | Wavelet Transforms |

**1. Introduction**

Application of data science in different disciplines is exponentially increasing. Because data science has had tremendous progresses in analysis and use of data. Like other disciplines, economics has benefited from the advancements of data science. Advancements of data science in economics have been progressive and have recorded promising results in the literature. The advances in data science have been remarkable in the three parts of deep learning, hybrid machine learnings, and ensemble machine learnings. Machine learning (ML) algorithms are able to learn from data with minimal human intervention. Deep Learning (DL) models, as a proper subset of machine learning models, are applied in many aspects of the today's society from the self-driving cares to image recognition and the earthquake prediction. Many evidences are provided in the literature that express the outperformance of DL models compared to the other machine learning methods such as SVM, KNN, GRNN in the fields of economics. The evolution of DL methods is very fast and every day, many sections and disciplines are added to the number of users and beneficiaries of DL algorithms. Hybrid machine learning models consist of two or more single algorithms to increase the accuracy of prediction [1]. Hybrid models can be formed by combining two predictive machine learning algorithms or a machine learning algorithm and an optimization method to maximize the prediction function [2]. It is demonstrated that the hybrid machine learning models outperform the single algorithms and such an approach has improved the predication accuracy [3-5]. Ensemble machine learning algorithms are one of the supervised learning algorithms that use multiple learning algorithms to improve learning processes and increase predictive accuracy [1]. Ensemble models apply different training algorithms to enhance training and learning from data.

The body of literature embraced the review papers on the state of the art of DL methods in different disciplines such as image recognition (e.g. [6]), animal behavior (e.g. [7]), renewable energy forecasting (e.g. [8]), and the review papers on hybrid methods in different various fields such as financial time series [9], solar radiation forecasting [10], and FOREX rate prediction [11], and the review papers on ensemble methods in fields such as breast cancer [12], image categorization[13], electric vehicle user behavior prediction [14], solar power generation forecasting [15].

Despite the fact that many researchers applied deep learning methods to address different problems in the field of Economics (see Figure 1), these studies are scattered. While, however, no single study exists to provide a comprehensive view of the contributions of data science in Economics related fields. Therefore, the current study is conducted to bridge this literature gap. In other words, the main objective of this study is to investigate the advancement of data science in the three parts of deep learning methods, hybrid deep learning methods, and finally ensemble machine learning techniques in Economics related fields. This study also classifies the sectors in which the such methods are applied.

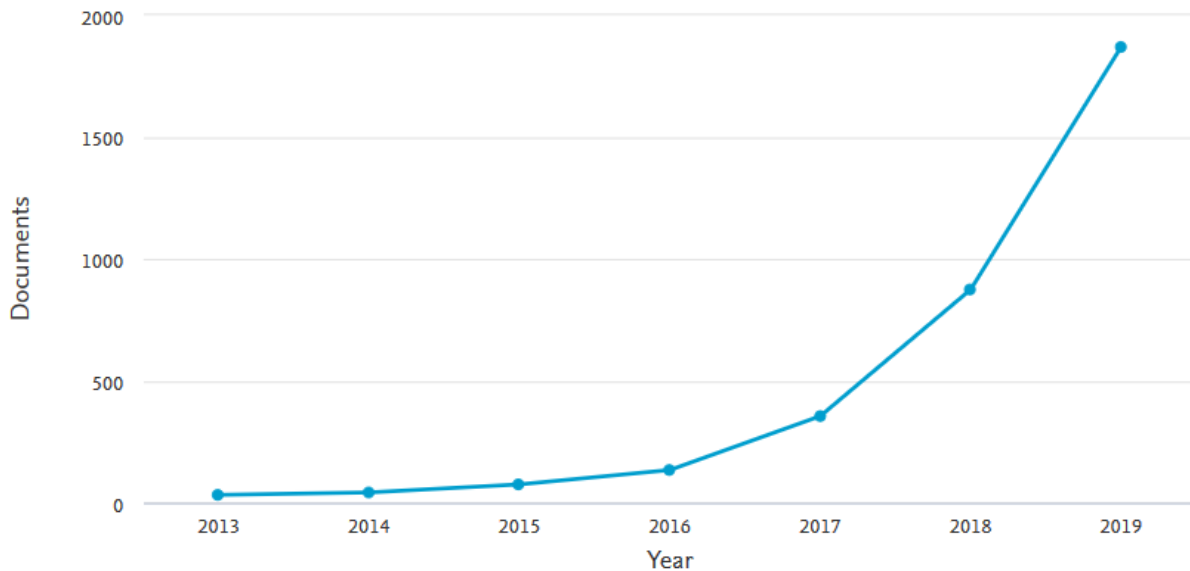

Figure 1: Increasing application of deep learning in Economics

In the following, first, the method by which the database of this article was formed is described. Subsequently, the articles were categorized by sector and area of study, and the models used in each category are described in detail. The study also classifies the models used in the reviewed articles, with each class being explained in detail. Finally, the findings of this study are analyzed, and the conclusions of the article are presented.

**2. Materials and Methods**

The current study applied Prisma, a systematic literature review approach, to find the most published articles applying DL methods for addressing an issue in a field related to Economics. Systematic literature review based on Prisma method includes four steps: 1) identification, 2) screening, 3) eligibility, 4) inclusion. In the identification stage, the documents are identified through initial search among the mentioned databases. In this study the review is limited to the original peer-review research articles published through Thomson Reuters Web-of-Science (WoS) and Elsevier Scopus. Besides, only articles written in English considered for the review in this study. This step resulted in 204 articles. The screening step includes two stages on which, first, duplicate articles are eliminated. As a result, 112 unique articles moved to the next stage, where the relevance of the articles is examined on the basis of their title and abstract. The result of this step was 69 articles for further consideration. The next step of Prisma model is eligibility, in which the full text of articles was read by the authors and 50 of them considered eligible for final review in this study. The last step of the Prisma model is the creation of the database of the study, which is used for qualitative and quantitative analysis. The database of the current study comprises 50 articles and all the analysis in this study took place based on these articles. Figure 2 illustrates the steps of creating the database of the current study based on the Prism method.

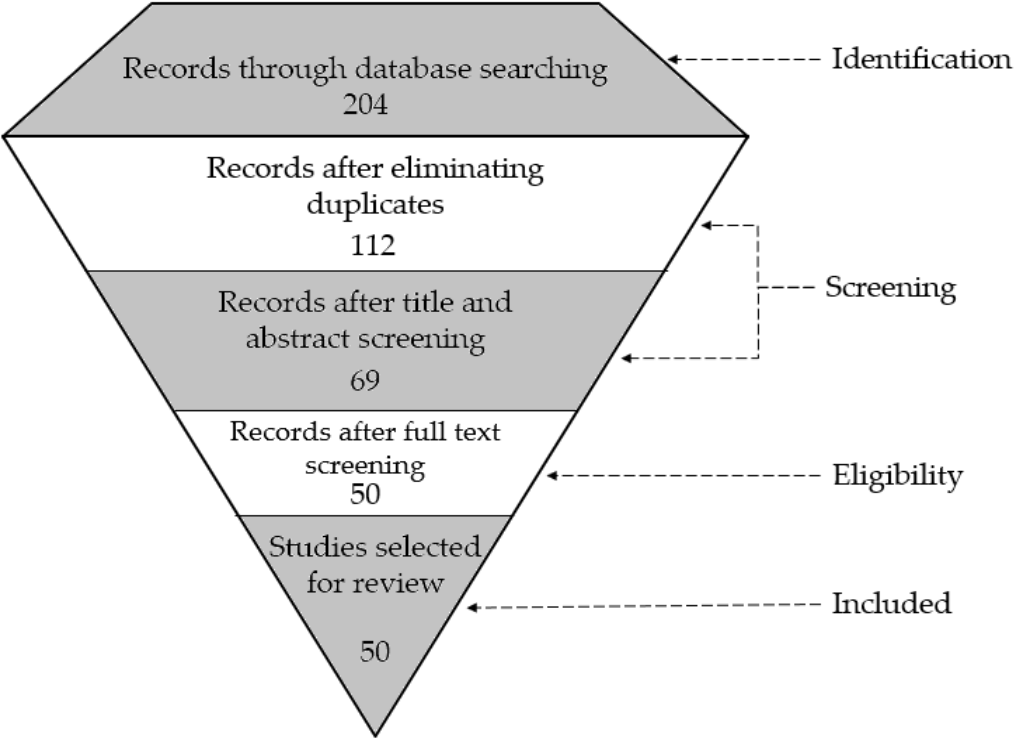

Figure 2. Diagram of systematic selection of the study database

## 3. Results

*3.1 Applications of deep learning models in the Economics*

In-depth review of 50 articles intended for review in this study identified four different areas in which DL was applied which were Stock market, marketing, E-commerce, corporate bankruptcy, and cryptocurrency. Reviewing the study database reveals that the use of hybrid deep neural networks has become very common in the study database. Hence, a section in this paper is devoted to the study of the applied hybrid deep neural network models as well.

3.1.1 Stock Market

Applying deep learning in the stock market has been more common than in other areas of economics as most of the research studies reviewed in the current study are classified in this category (33 out of 50). Table 2 presents various motivations stimulated researchers to use predictive models in stock market studies. Investment in the stock market is profitable, while higher the profit, higher the risk. Therefore, the investors always try to determine and estimate the stock value before any action. The stock value is often influenced by a sort of uncontrollable economic and political factors that make it notoriously difficult to identify the future stock market trends. On the one hand, the nature of stock market is so volatile and complex, and on the other hand, the financial time series data are so noisy and nonstationary. Thus, the traditional forecasting models are not reliable enough to predict the stock value and researchers are seeking new methodologies based on DL models to enhance the accuracy of such predictions. Forecasting stock value has been the objective of 25 out of 33 articles. In addition, there are studies aimed at applying DL for the purpose of sentiment analysis (sentiment analysis refers to analyzing the context of texts to affectively extracts subjective information) to find the future trends in the stock market. Besides, portfolio management, algorithmic trading (using a pre-programmed automated system for trading), automated stock trading, socially responsible investment portfolios, the S&P 500 index trend prediction, exchange-trade-fund (EFT) options prices prediction were the objectives of other articles stimulating to employ DL methods. Financial time series have been the data source of all these studies, except for the studies aimed at sentiment analysis. In other words, these studies applied different DL models to find the patterns among the financial time series.

LSTM

Long short term memory (LSTM) networks are a special kind of recurrent neural networks (RNN), which can overcome the main issue of RNN, i.e., vanishing gradients by using the gates to selectively retain the relevant information and discard the unrelated information. The structure of an LSTM neural network is shown in Fig. 1, which is composed of a memory unit $C$, a hidden state $h$ and three types of gates. Specifically, for each step $t$, LSTM receives an input $x_t$ and the previous hidden state $h_{t-1}$. Then, it calculates activations of the gates. Finally, the memory unit and the hidden state are updated. The computations involved are described below:

$$f_t = \sigma(W_f x_t + w_f h_{t-1} + b_f)$$

$$i_t = \sigma(W_i x_t + w_i h_{t-1} + b_i)$$

$$O_t = \sigma(W_o x_t + w_o h_{t-1} + b_o)$$

$$C_t = f_t \text{ e } C_{t-1} + i_t \text{ e } \sigma_c(W_c x_t + w_c h_{t-1} + b_c)$$

$$h_t = O_t \text{ e } \tanh(C_t)$$

Where $W$, $w$ and $b$ denote the weights of input, weights of recurrent output, and biases. $f$, $i$ and $O$ represent the forget, input, and output gate vectors, respectively. e is the element-wise multiplication.

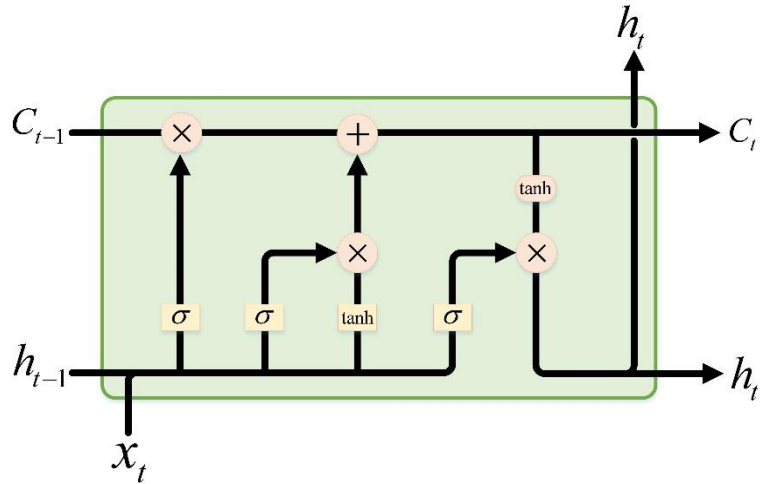

Figure 3 The structure of the LSTM network

Relying on LSTM algorithm, either a single long short time memory (LSTM) or a hybrid model of LSTM, many researchers strived to forecast the stock value. Adapting LSTM algorithm, Moon and Kim [16] propose an algorithm to predict the stock market index and the stock market volatility. Fischer and Krauss [17] expand the LSTM networks to forecast out-of-sample directional movements in the stock market. The comparison between the performance of their model with the RF, DNN, LOG illustrates a remarkable outperformance of the LSTM model. Tamura et al. [18] introduce a two-dimensional approach to predict the stock values in which the financial technical indexes of Japanese stoke market entered as input data to the LSTM for the prediction and then the data on financial statements of the related companies are retrieve and added to the database. Wang et al. [19] tried to find the best model to predict financial time-series for portfolio management to optimize the portfolio formation. They compare the results of the LSTM against SVM, RF, DNN, and ARIMA and they realized that LSTM is more suitable for financial time-series forecasting in their study. Using LSTM, Fister et al. [20] design a model for automated stock trading. They argue that the performance of LSTM is remarkably higher than the traditional trading strategies such as passive and rule-based trading strategies, in their case studies the German blue-chip stock and BMW in the period between 2010 and 2018.

In addition, there are many evidences in the literature that hybrid LSTM methods also outperform the other single DL methods. In the application of stock market, LSTM is hybrid with different methods. Tamura et al. [18], for instance, report that the results of the accuracy test outperform the model in the literature, in their case study. Employing O-LSTM, Agrawal et al. [21] propose a model for the stock price prediction. Agrawal et al. [21] use correlation-Tensor which is formed by Stock Technical Indicators to optimize the deep learning function. Agrawal et al. [21] integrate optimized LSTM with STIs and develop two predictive models one for price trend prediction and the other for taking Buy-Sell decision at the end of day. Integrating WT, SAEs and LSTM, Bao, Yue, and Rao [22] propose a new method to predict the stock price. According to Bao et al. [22], WT, firstly, eliminate noises to decompose the stock price time series. In the next stage, predictive features for the stock price are created by SAEs. And finally, the LSTM is applied to predict the next day's closing price based on the features generated through the previous stage. Bao et al. [22] claim that their model outperforms the state-of-the-art models in the literature in both in terms of predictive accuracy and profitability performance. To cope with non-linearity and non-stationary characteristics of financial time series, Yan and Ouyang [23] integrate WA-LSTM to forecast the daily closing price of the Shanghai Composite Index. Their results show that their proposed model outperformed MLP, SVM and KNN in finding the patterns in the financial time series. Vo et al. [24] use a MB-LSTM to develop a Deep Responsible Investment Portfolio (DRIP) model for the prediction of stock returns for socially responsible investment portfolios. They applied the DRL model to retrain neural networks. Fang, Chen, Xue [25] develop a methodology to predict the exchange-trade-fund (EFT) options prices. Integrating LSTM model and SVR, They firstly develop two models of LSTM-

SVR I and LSTM-SVR II where in LSTM-SVR I the output of LSTM and the final transaction price, buy price, highest price, lowest price, volume, historical volatility, and the implied volatility of the time segment, that considered as factors affecting the price, added as the input of SVR model. Whilst, in LSTM-SVR II, the hidden state vectors of LSTM and the seven factors affecting the option price considered as the SVR's inputs. They also compare the results with the LSTM model and the RF model.

Table 1. Application of data science algorithms in Stock Market

| Row | Modeling Methods | Data Source | Research Objective | Source |
|---|---|---|---|---|
| 1 | LSTM Comparing with SVM, RF, DNN, and ARIMAs | Financial Time Series | Portfolio management | Wang et al. [19] |
| 2 | time-driven feature-aware and DRL | Financial Time Series | Algorithmic trading | Lei et al. [26] |
| 3 | Multivariate Bidirectional LSTM Comparing with DRL | Financial Time Series | Socially Responsible Investment Portfolios | Vo et al. [24] |
| 4 | GRU – CNN | Financial Time Series | Stock Price Prediction | Sabeena and Venkata Subba Reddy [27] |
| 5 | Adam optimizer-MDNN | Financial Time Series | Stock Price Prediction | Das and Mishra [28] |
| 6 | DNN | Financial Time Series | Stock Price Prediction | Go and Hong [29] |
| 7 | O-LSTM-STI | Financial Time Series | Stock Price Prediction | Agrawal et al. [21] |
| 8 | CNN comparing with DNNC and LSTM | Financial Time Series | Stock Price Prediction | Gonçalves et al. [30] |
| 9 | DNN-SLR | Financial Time Series | Stock Price Prediction | Moews et al. [31] |
| 10 | DNN | Financial Time Series | Stock Price Prediction | Song et al. [32] |
| 11 | LSTM-SVR comparing with RF and LSTM | Financial Time Series | Exchange-trade-fund (EFT) Options Prices Prediction | Fang et al. [25] |
| 12 | MFNN (CNN and RNN) | Financial Time Series | Stock Price Prediction | Long et al. [33] |
| 13 | LSTM | Financial Time Series | Automated Stock Trading | Fister et al. [20] |
| 14 | Random Forest, SVM and K-neighbors | Financial Time Series | Stock Price Prediction | Rajesh [34] |
| 15 | LSTM | Financial Time Series | Stock Price Prediction | Moon and Kim [16] |
| 16 | CNN comparing with ANN and SVM | Financial Time Series | Stock Price Prediction | Sim Kim, and Ahn [35] |
| 17 | LSTM-STIs | Financial Time Series | Stock Price Prediction | Agrawal et al. [36] |
| 18 | CNN | Financial Time Series | Stock Price Prediction | Tashiro et al. [37] |
| 19 | LSDL | Financial Time Series | Stock Price Prediction | Sirignano and Cont [38] |
| 20 | BRT comparing with NNRE, SVRE, and RFR | Financial Time Series | Stock Price Prediction | Weng et al. [39] |
| 21 | ELM-AE comparing with GARCH, GRNN, MLP, RF and GRDH | Financial Time Series | Stock Price Prediction | Preeti et al. [40] |
| 22 | CNN comparing with doc2vec and LSTM | Social media | Sentiment Analysis | Sohangir et al. [41] |
| 23 | LSTM comparing with RF, DNN and LOG | Financial Time Series | Stock Price Prediction | Fischer and Krauss [17] |
| 24 | two-stream GRU | Financial news | Sentiment Analysis | Lien Minh et al. [42] |
| 25 | DNN | Financial Time Series | The S&P 500 Index Trend Prediction | Das et al. [43] |
| 26 | wavelet analysis with LSTM, comparing with SVM, KNN and MLP | Financial Time Series | Stock Price Prediction | Yan and Ouyang [23] |
| 27 | MACN | Financial Time Series | Stock Price Prediction | Kim et al. [44] |
| 28 | EL-ANN | Financial Time Series | Stock Price Prediction | Faghihi-Nezhad and Minaei-Bidgoli [45] |
| 29 | LSTM | Financial Time Series | Stock Price Prediction | Tamura et al. [18] |
| 30 | DNN comparing with PCA, Autoencoder, and RBM | Financial Time Series | Stock Price Prediction | Chong et al. [46] |
| 31 | CNN | Financial Time Series | Stock Price Prediction | Dingli and Fournier [47] |
| 32 | (2D)2PCA – DNN comparing with RBFNN | Financial Time Series | Stock Price Prediction | Singh and Srivastava [48] |
| 33 | WT-SAEs-LSTM | Financial Time Series | Stock Price Prediction | Bao et al. [22] |

DNN

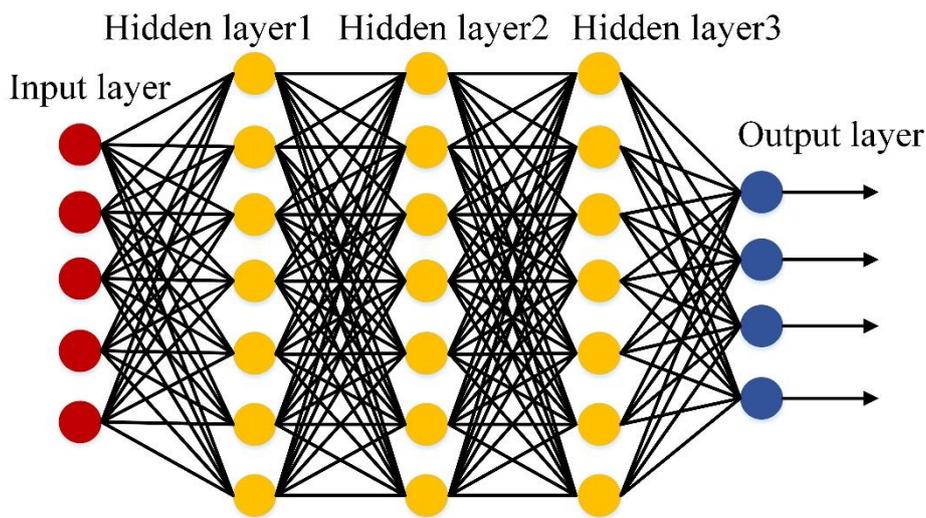

Figure 4 Structure of the deep neural network

Deep neural network (DNN) is the other DL method, which is composed of multiple levels of nonlinear operations, and each layer only receives the connections from its previous training layer

(as shown in Figure 4). Suppose $X$ be the input data, and $w_j$ be a filter bank. The multi-layer features of the DNN can be computed as:

$$f(X) = f_k(\text{L } f_2(f_1(X; w_1); w_2)\text{L }, w_k).$$

Currently, DNN has widely applied in the stock market to identify the trends and patterns among the financial time series data. Go and Hong [29] applied DNN method to predict the stock value. They firstly train the method by the time series data and then test and confirm the predictability of their model. Song et al. [32] develop DNN using 715 novel input-features to forecast the stock price fluctuation. They also compare the performance of their model with the other models that include simple price-based input-features. For predicting the stock market behavior, Chong, Han, and Park [46] examine the performance of DNN. They consider high-frequency intraday stock returns as the input in their model. They analyze the predictability of PCA, autoencoder, and RBM. According to their results, DNN have good predictability with the information they receive from the residuals of the autoregressive mode. Whilst, applying the autoregressive model to the residuals of the network cannot contribute to the predictability of the model. In addition, Chong et al. [46] found out applying covariance-based market structure analysis to the predictive network remarkably increase the covariance estimation. Das et al. [43] use DNN to predict the future trends of the S&P 500 Index. Their results show that their model can poorly forecast the behavior of the underlying stocks in the S&P 500 index. They believe that randomness and non-stationarity are the reasons making hard the predictability of this index.

Besides, hybrid methods that are constructed based on DNN have been reported very accurate in the financial time series data. Das and Mishra [28], for instance, propose an advanced model to plan and analyze and predict the stock value. They use a MDNN optimized by Adam optimizer to find the patterns among the stock values. Moews et al. [31] propose a method to predict the behavior of the stock market, as a complex system with a massive number of noisy time series. Their model integrates DNN and SLR. Moews et al. [31] consider regression slopes as trend strength indicators for a given time interval. To predict the Google stock price, Singh and Srivastava [48] compare two integrated models (2D)2PCA-DNN and (2D)2PCA-RBFNN. According to their results, (2D)2PCA-DNN model has a higher accuracy in predicting the stock price in their case study. They also compare their results with RNN model, and they reported that the predictability (2D)2PCA-DNN outperforms RNN.

CNN

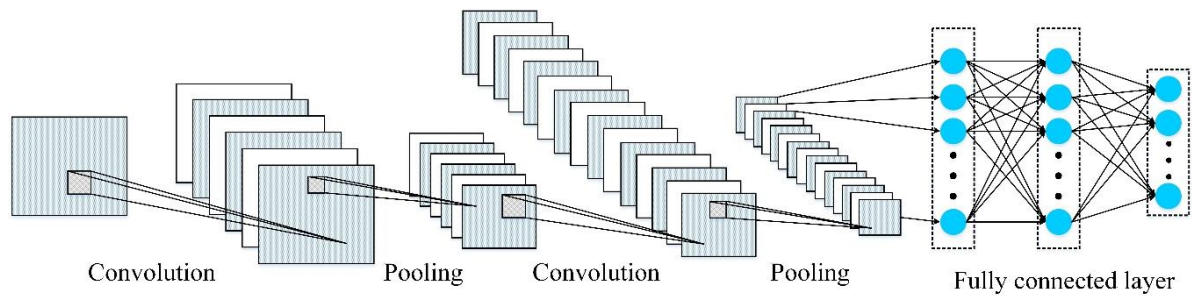

Figure 5 Structure of the CNN

Conventional neural network (CNN) is one of the most popular methods in deep learning, which is widely applied in various fields [49-52], such as classification, language processing, and object detection. A classical structure of CNN is presented in Figure 5, which mainly consists of three components, i.e., convolution layers, pooling layers, and fully connected layers. Different layers have different roles in the training process. Here, we will briefly review those layers:

1) Convolutional layer: Convolutional layer is composed of a set of trainable filters, which is used

to perform feature extraction. Suppose $X$ be the input data. There are $k$ filters in convolutional layers, and thus, the output of the convolutional layer can be denotes as follows:

$$y_j = \sum_i f(x_i * w_j + b_j), j = 1, 2, ..., k$$

where $w$ and $b$ are the weight and bias, respectively. $f(\cdot)$ denotes an activation function. $*$ represents the convolution operation.

2) Pooling layer: In general, the pooling layer is used to decrease the dimensions of the obtained feature data and network parameters. Currently, max pooling and average pooling are the most widely used ways. Let $S$ be a $p \times p$ window size, the average pooling operation can be expressed as follows:

$$z = \frac{1}{N} \sum_{(i,j) \in S} x_{ij}$$

Where $x_{ij}$ indicates the activation value at $(i, j)$. $N$ denotes the total number of elements in $S$.

3) Fully connected layer: Following the last pooling layer, the fully connected layer is utilized to reshape the feature maps into 1-D feature vector, which can be expressed as:

$$Y = \sum_i f(WZ + b)$$

Where $Y$ and $Z$ denote the output vector and the input features. $W$ and $b$ represents the weight and bias of fully connected layer.

Recently, CNN has extensively applied by many researchers for predicting the stock values using the financial time series data. Sim et al. [35] develop a model to predict the stock price. They adapt CNN to develop their model. In their study, the result of comparing the predictive performance of CNN with ANN and SVM illustrated that the CNN proved a better performance in forecasting stock price. Tashiro et al. [37] firstly criticize the current models for the price prediction in the stock markets that these models ignore properties of market orders. Therefore, they come up with the CNN architecture integrating order-based features to predict the mid-price trends in the stock markets. Their results prove that adding the features of orders to the model has increased the accuracy of the model. Dingli and Fournier [47] applied the CNN to predict the future movement of stock prices. They report that the predictive accuracy of their model was 65% when they study the next month price and 60% when they look for the next week price. Gonçalves et al. [30] compare the results of prediction of CNN, LSTM and DNNC for finding the best model to predict the price trends in the exchange markets. Their finding reveals that CNN, on average, has the best predictive power in their case studies. Sohangir et al. [41] compare the performance of several neural network models like CNN, LSTM, and doc2vec for sentiment analysis among the posts and opinions of the experts in StockTwits to predict the movements in the stock markets. Their results disclose that a CNN has had the highest accuracy to predict sentiment of experts in their case study. Paolanti et al. [53] employ deep CNN to develop a mobile robot, so called ROCKy, to analyze real-time store heat maps of retail store shelves for detection of Shelf Out of Stock (SOOS) and Promotional Activities, during working hours.

To increase the accuracy of CNN, some researchers integrated CNN with other models and have proposed a new hybrid model. For example, integrating GRU and CNN, Sabeena et al. [27] introduce a hybrid DL model to predict financial fluctuations in the real-time stock market that is able to process the real-time data from online financial sites. To predict the price movement from financial time series samples, Long et al. [33] introduce an end-to-end model called MFNN. They incorporate CNN and RNN to construct the multi-filters structure of their model.

Other algorithms

In addition to LSTM, DNN, and CNN other DL methods are employed for prediction of stock value using time series data. For example, Sirignano and Cont [38] develop a LSDL model to study the USA market quotes and transactions. Their results disclose that there is a universal and stationary relationship between order flow history and the price trends. Kim et al. [44] develop MACN model to optimize financial time series data. They claim that the contribution of their mode is that this model is able to share and generalize the experience by agent in the stock trading.

Besides, various other hybrid methods are applied by the researchers for financial time series data. For example, to predict the stock prices trends, Lien Minh et al. [42] develop the two-stream GRUN and Stock2Vec model. They believe that financial news and sentiment dictionary affect the stock prices therefore, their method trained to analyze the sentiments in the financial news and their relationship with the financial prices. Their finding supported the outperformance of their model in compare with the current models. Lien Minh et al. [42] also claim that Stock2Vec is very efficient in financial datasets. Lei et al. [26], by deep learning model and reinforcement learning model, develop a time-driven feature-aware jointly deep reinforcement learning model (TFJ-DRL) for financial time-series forecasting in algorithmic trading. Preeti et al. [40] introduce an ELM-AE model to find the patterns in the financial time series. They test the accuracy of their model by the time series data of Gold Price and Crude Oil Price and also compare the results with those of GARCH, GRNN, MLP, RF and GRDH. The result of mean square error (MSE) test proved that the performance of their model was higher than existing methods.

Rajesh et al. [34] use ensemble learning to predict the future stock trend. They applied Heat Map and Ensemble Learning to financial data of top 500 companies in S&P stock exchange. Their results show that the combination of Random Forest, SVM and K-neighbors classifiers had the most accurate results and the accuracy of the proposed model was 23% higher than a single classifier labelling prediction model. Weng et al. [39] tried to design a financial expert system to forecast short term stock prices. To analyze the collected data and predict the stock prices, they employed four machine learning ensemble methods of neural network regression ensemble, support vector regression ensemble, boosted regression tree, and random forest regression. They, using Citi Group stock ($C) data, forecast the 1-day ahead price of 19 stocks from different industries. Weng et al. [39] claimed that boosted regression tree outperformed other ensemble models with a considerable mean absolute percent error (MAPE) that was better than what was reported in the literature. Faghihi-Nezhad and Minaei-Bidgoli [45] using ensemble learning and ANN proposed a two-stage model to predict the stock price. In their propose model firstly the direction of the next price movement is predicted and then they create a new training dataset to forecast the stock price. They use genetic algorithm (GA) optimization and particle swarm optimization (PSO) to optimize the results of each stage. They claimed that the accuracy of their model in prediction of stock price outperformed other models in the literature.

A close look at the reviewed articles that categorized in stock market category in the current study reveals that although the research objectives of the articles are different, they mainly utilize financial time series data (i.e. 31 out of 33 articles) and only 2 article have used the financial news and social media as the data source to find the future trends in the stock market (see Table 3).

Table 2. Classification of Articles Using Data Science by Research Purpose and Data Source in the Stock Market Section

| Research Objective | Data Source | Number of documents |
| --- | --- | --- |
| Stock Price Prediction | Financial Time Series | 25 |
| Sentiment Analysis | Financial News, Social Media | 2 |
| Portfolio management | Financial Time Series | 1 |
| Algorithmic trading | Financial Time Series | 1 |
| Socially Responsible Investment Portfolios | Financial Time Series | 1 |
| Automated Stock Trading | Financial Time Series | 1 |
| The S&P 500 Index Trend Prediction | Financial Time Series | 1 |
| Exchange-trade-fund (EFT) Options Prices Prediction | Financial Time Series | 1 |

3.1.2 Marketing

Studying the purpose of the articles reviewed in the current study disclosed that the authors applied DL and methods for studying customer behavior. That is why these articles are classified in a group labeled marketing. As it is presented in Table 4, 2 of the studies have applied a single DL method and three of them used a hybrid DL method. In addition, these studies use various data sources such as customer time series data, case study, and social media. For example, Paolanti et al. [53] employ DCNN to develop a mobile robot, so called ROCKy, to analyze real-time store heat maps of retail store shelves for detection of Shelf Out of Stock (SOOS) and Promotional Activities, during working hours. Dingli, Marmara, and Fournier [54] were looking for solutions to find the patterns and features among transactional data to predict customer churn within the retail industry. To do so, they compare the performance of CNN and RBM and they found out the RBM attained outperformed in customer churn prediction.

Table 3. Application of data science algorithms in Marketing

| Application | Modeling Methods | Data Source | Research Objective | Source |
|---|---|---|---|---|
| Marketing | RF – DNN | Time Series data of Customers | Customer Behavior | Ładyżyński et al. [55] |
| | RF | Time Series data of Customers | Customer Behavior | Ullah et al. [56] |
| | DCNN | Primary Data | Detection of Shelf Out of Stock (SOOS) and Promotional Activities | Paolanti et al. [53] |
| | RNNs- CNNs | Social media | Sentiment Analysis | Agarwal [57] |
| | SN-CFM | Social media | Customer behavior | Ardabili et al. [58] |
| | RBM | Primary Data | Customer behavior | Dingli et al. [54] |

On the other hand, RF-DNN, RNN-CNN and SN-CFM are hybrid models that researchers proposed to study the customer behavior. Ładyżyński et al. [55], for instance, using Random Forest (RF) and DNN methods and customers' historical transactional data propose a hybrid model that is able to predict the customers' willingness to purchase credit products from the banks. Ullah et al. [56] use the RF algorithm to prediction churn customers and use their result to formulate strategies for customer relationship management to prevent churners. Ullah et al. [56] explain that combination of churn classification utilizing the RF algorithm and customer profiling using k-means clustering increased their model performance. Agarwal [57] integrated RNN and CNN to develop a model for sentiment analysis. According to Agarwal [57], sentiment analysis is the best approach to get the customers' feedback. He tested their model using the social media data and believe that the result of the sentiment analysis provides guidance to the business to improve the quality of their service and presents evidences for the startups to improve customer experience. Shamshoddin et al. [58] propose SN-CFM to predict the consumer preferences according to similarity of features of users and products that are acquired from the internet of things and social networks.

3.1.3 E-commerce

Another category emerged after reviewing the articles is labeled E-commerce where the authors employed data science models to address problems in E-commerce sector. A summary of these studies is presented in Table 5. Lei [59], based on the GRU model, designs a neural network investment quality evaluation model to support the decision-making related to investment in e-commerce. According to Lei [59], their proposed model is able to evaluate different index data that provide a better picture to investors. Cai et al. [60] use deep reinforcement learning to develop an algorithm to address allocating impression problems in E-commerce websites, such as Taobao, eBay, and Amazon. In this algorithm the buyers are allocated to the sellers based on the buyer's impressions and the seller's strategies in a way that to maximize the income of the platform. To do so they applied GRU and their finding shows that GRU outperform the deep "Deterministic Policy Gradient (DDPG). Ha, Pyo, Kim [61] apply RNN to develop deep categorization network (DeepCN) for item categorization in E-commerce. Item categorization refers to classifying the leaf category of items from their metadata. They use RNN to generate features from text metadata and categorizing the items accordingly. Wang, Mo, and Tseng. [62] use RNN to develop a personalized product recommendation system in E-commerce websites. The result of testing their model discloses the outperformance of RNN to K nearest neighbors (KNN). Wu and Yan [63] claim that the main assumption of the current production recommender models for E-commerce websites is that all the historical data of users are recorded while in practice many platforms fail to record such dada. Therefore, they come up with a list-wise DNN to model the temporal online user behaviors and provide recommendations for the anonymous users.

Table 4. Application of data science algorithms in E-commerce

| Application | Modeling Methods | Data Source | Research Objective | Source |
|---|---|---|---|---|
| E-commerce | GRU | Financial Time Series | Investment Quality Evaluation Model | Lei [59] |
| | GRU | Customers Time Series | Impression Allocation Problem | Cai et al. [60] |
| | RNN | Primary Data | Item Categorization | Ha et al. [61] |
| | RNN | Primary Data | Product Recommendation | Wang et al. [62] |
| | LWDNN | Customers Time Series | Product Recommendation | Wu and Yan [63] |

3.1.4 Cryptocurrency

The decision-making process related to investing on the cryptocurrencies is similar to investing to the stock market where the prediction of future value is very determinant and effective on the investment decisions. Applying machine learning and DL models to predict the trends of cryptocurrencies' prices is an attractive research problem which is emerging in the literature (see Table 6). Lahmiri and Bekiros [64], for example, applied deep learning methods for the prediction of price of cryptocurrencies including Bitcoin, Digital Cash and Ripple. They compare the predictive performance of LSTM and GRNN. Their finding discloses that the LSTM model has a better performance in their case studies in compare with GRNN. Altan, Karasu, and Bekiros [65] claim that integrating LSTM and empirical wavelet transform (EWT) improve the performance of LSTM in forecasting the digital currency price. In this study, Altan et al. (2019) test their proposed model using the Bitcoin, Ripple, Digital Cash and Litecoin time series data. Jiang and Liang [66] develop a CNN model to predict the price of Bitcoin as a cryptocurrency example. They train their proposed model by historical data of financial assets prices and output of their model was designed to be portfolio weights of the set.

Table 5. Application of data science algorithms in Cryptocurrency

| Application | Modeling Methods | Data Source | Research Objective | Source |
|---|---|---|---|---|
| Cryptocurrency | LSTM comparing with GRNN | Financial Time Series | Cryptocurrencies Price prediction | Lahmiri and Bekiros [64] |
| | LSTM-EWT | Financial Time Series | Cryptocurrencies Price prediction | Altana et al. [65] |
| | CNN | Financial Time Series | Cryptocurrencies Price prediction | Jiang and Liang [66] |

3.1.5 Corporate bankruptcy prediction

Corporate bankruptcy prediction has become an important tool to evaluate the future financial situation of the companies. Utilizing machine learning based methods is widely recommended to address bankruptcy prediction problems. To address corporate bankruptcy prediction, Chen, Chen, Shi [67] utilize bagging and boosting ensemble strategies and develop two models of Bagged-pSVM and Boosted-pSVM. Using datasets of UCI and LibSVM, they test their models and explain that ensemble learning strategies increased the performance of the models in of bankruptcy prediction. Lin, Lu, and Tsai [68] believe that finding the best match of feature selection and classification techniques improves the prediction performance of bankruptcy prediction models. Their results reveal that the genetic algorithm as the wrapper-based feature selection method and the combination of the genetic algorithm with the naïve Bayes and support vector machine classifiers had a remarkable predictive performance. Lahmiri et al. [69], to develop an accurate model for forecasting corporate bankruptcy, compare the performance of different ensemble systems of AdaBoost, LogitBoost, RUSBoost, subspace, and bagging. Their finding reveals that that AdaBoost model has been effective in terms of short time data processing and low classification error, and limited complexity. Faris et al. [70] investigate the combination of re-sampling (oversampling) techniques and multiple features election methods to improve the accuracy of bankruptcy prediction methods. According to their results, employing SMOTE oversampling technique and AdaBoost ensemble method using REP tree provides reliable promising results to bankruptcy prediction. A summary of these research articles is presented in Table 7.

Table 6. Application of data science algorithms in corporate bankruptcy prediction

| Application | Modeling Methods | Data Source | Source |
|---|---|---|---|
| Bankruptcy Prediction | Bagged-pSVM and Boosted-pSVM | UCI and LibSVM datasets | Chen et al. [67] |
| | Genetic Algorithm with the Naïve Bayes and SVM classifiers | Australian credit, German credit, and Taiwan bankruptcy datasets | Lin et al. [68] |
| | AdaBoost | University of | Lahmiri et al. [69] |

| | | California Irvine (UCI) Machine Learning Repository | |
|---|---|---|---|
| | SMOTE- AdaBoost- REP Tree | Infotel database | Faris et al. [70] |

*3.2. Deep learning methods*

DL models are structured based on the artificial neural network and the modern algorithms of deep learning rooted in the work of Hinton et al. [71]. Hinton et al. [71] develop, for the first time, a two-step approach in which the deep learning algorithms are firstly trained and then fine-tune the model in a back-through process. The advantage of DL models compared to other ML models is that DL models can effectively identify high-level features and representations (the outputs) from a large diverse data sample (inputs). Employing an unsupervised pre-training and a supervised fine-tuning approach, the deep learning models extract hierarchical features from the inputs to classify the patterns of the data [72]. The ability of DL algorithms in prediction and finding the patterns among the raw data has grabbed the attraction of many researchers from various fields. Economics researchers have applied DL models for a variety of reasons such as stock price prediction (e.g. [16-18]) and forecasting consumer behavior (e.g.[54]), etc. LSTM, CNN, and DNN are respectively the most applied DL models among the database of the study. LSTM is applied to stock price prediction [16-18], portfolio management [19], automated stock trading [20], and cryptocurrencies price prediction [64]. Interestingly enough, LSTM method has only applied to find the patterns among financial time series data.

Similar to LSTM, the CNN algorithm is applied mainly for financial time series data to stock price prediction [30, 35, 37, 47] and cryptocurrencies price prediction [66]. CNN algorithm is also used for analyzing social media data for the purpose of sentiment analysis [41]. DNN algorithm, likewise LSTM, has only applied for analyzing financial time series data to predict the stock price [29, 32, 46] and the S&P 500 Index trend prediction [43]. GRU algorithm applied in the E-commerce section to both analyze financial time series [59] and customers time series [60]. RNN is a DL algorithm applied to analyze primary data for item categorization [61] and product recommendation [62]. LSDL and MACN are DL algorithms that are used to analyze financial time series data to predict stock price [38, 44]. Ultimately, it is found that DCNN and RBM applied for analyzing primary data for respectively Promotional Activities [57] and Customer behavior forecasting [54] (see Table 8).

Table 7. List of single deep learning methods employed in Economics related fields

| Method | Applications | | | | Frequency | References |
|---|---|---|---|---|---|---|
| | Stock Market | Marketing | Cryptocurrency | E-Commerce | | |
| LSTM | X | - | X | - | 6 | Moon and Kim [16], Fischer and Krauss [17], Tamura et al. [18], Wang et al. [19], Fister et al. [20], Lahmiri and Bekiros [64] |
| CNN | X | - | X | - | 6 | Gonçalves et al. [30], Sim Kim, and Ahn [35], Tashiro et al. [37], Sohangir et al. [41], Dingli |
| | | | | | | and Fournier [47], Jiang and Liang [66] |
| DNN | X | - | - | - | 4 | Go and Hong [29], Song et al. [32], Das et al. [43], Chong et al. [46] |
| GRU | - | - | - | X | 2 | Lei [59], Cai et al. [60] |
| RNN | - | - | - | X | 2 | Ha et al. [61], Wang et al. [62] |
| LSDL | X | - | - | - | 1 | Sirignano and Cont [38] |
| MACN | X | - | - | - | 1 | Kim et al. [44] |
| DCNN | - | X | - | - | 1 | Paolanti et al. [53] |
| RBM | - | X | - | - | 1 | Dingli et al. [54] |

*3.3. Hybrid deep learning methods*

Hybrid deep neural networks are architectures that apply generative and discriminative components at the same time. Hybrid models combines machine learning models or combine a machine learning model with an optimization model to improve the predictivity of the machine learning model [3]. The findings of the literature review in the study reveal that hybrid deep learning (HDL) models are widely applied in the field of Economics. On the other hand, Figure 4 illustrates that the accuracy of such models is reported higher than the single DL models. The various HDL models used among the papers reviewed in this study are summarized in Table 9.

Table 8. List of hybrid deep learning models employed in Economic related fields

| Application | The hybrid method | Source |
|---|---|---|
| Stock Market | TDFA-DRL | Lei et al. [26] |
| | MB-LSTM | Vo et al. [24] |
| | GRU – CNN | Sabeena and Venkata Subba Reddy [27] |
| | AO-MDNN | Das and Mishra [28] |
| | OLSTM-STI | Agrawal et al. [21] |
| | DNN-SLR | Moews et al. [31] |
| | LSTM-SVR | Fang et al. [25] |
| | MFNN (CNN and RNN) | Long et al. [33] |
| | LSTM-STIs | Agrawal et al. [36] |
| | ELM-AE | Preeti et al. [40] |
| | TS-GRU | Lien Minh et al. [42] |
| | WA-LSTM | Yan and Ouyang [23] |
| | (2D)2PCA – DNN | Singh and Srivastava [48] |
| | WT-SAEs-LSTM | Bao et al. [22] |
| | RNNs- CNNs | Agarwal [57] |
| | SN-CFM | Shamshoddin, et al. [58] |
| E-commerce | LWDNN | Wu and Yan [63] |
| Cryptocurrency | LSTM-EWT | Altan et al. [65] |

*3.4. Ensemble machine learning algorithms*

Ensemble machine learning algorithms use multiple learning algorithms to improve training processes and boost learning from data [1]. There are many evidences in the literature that ensemble classifiers and ensemble learning systems have been very effective in financial time series data. For instance, Rajesh et al. [34] combine of RF, SVM and K-neighbors classifiers, Weng et al. [39] use neural network regression ensemble, and Faghihi-Nezhad and Minaei-Bidgoli [45] apply a ANN-EL to predict stock values. Besides, Chen et al. [67], Lin et al. [68], Lahmiri et al. [69] respectively applied Bagged-pSVM and Boosted-pSVM models, genetic algorithm with the naïve Bayes and SVM, SMOTE- AdaBoost- REP Tree to predict corporate bankruptcy. Ullah et al. [56] also take advantage of ensemble learning strategies and use a RF algorithm to predict the churn customers. Table 10 summers these studies which have used an ensemble model.

Table 9. List of ensemble models applied in the database on the current study

| Application | The ensemble method | Source |
| --- | --- | --- |
| Stock Market | RF-SVM-K-neighbors | Rajesh et al. [34] |
| | NNRE | Weng et al. [39] |
| | ANN-EL | Faghihi-Nezhad and Minaei-Bidgoli [45] |
| Corporate Bankruptcy | Bagged-pSVM and Boosted-pSVM | Chen et al. [67] |
| | Genetic Algorithm with the Naïve Bayes and SVM | Lin et al. [68] |
| | SMOTE- AdaBoost- REP Tree | Lahmiri et al. [69] |
| Marketing | RF – DNN | Ładyżyński et al. [55] |
| | RF | Ullah et al. [56] |

## 4. Discussion

Findings of this article disclosed that advancements of data science have contributed in four different areas of Economics discipline that are stock market, marketing, corporate bankruptcy, E-commerce, and cryptocurrency. It revealed that data science algorithms mainly applied for financial time series data to forecast stock prices in which LSTM model has been the most popular model for analyzing financial time series. CNN and DNN have been respectively the most applied algorithms among the reviewed articles in this study. Overall, 35 unique algorithms employed among 50 reviewed articles (see Figure 3) in which 9 of them used 9 single DL models (see Table 8), 18 HDL models (see Table 9), and 8 ensemble models (see table 10).

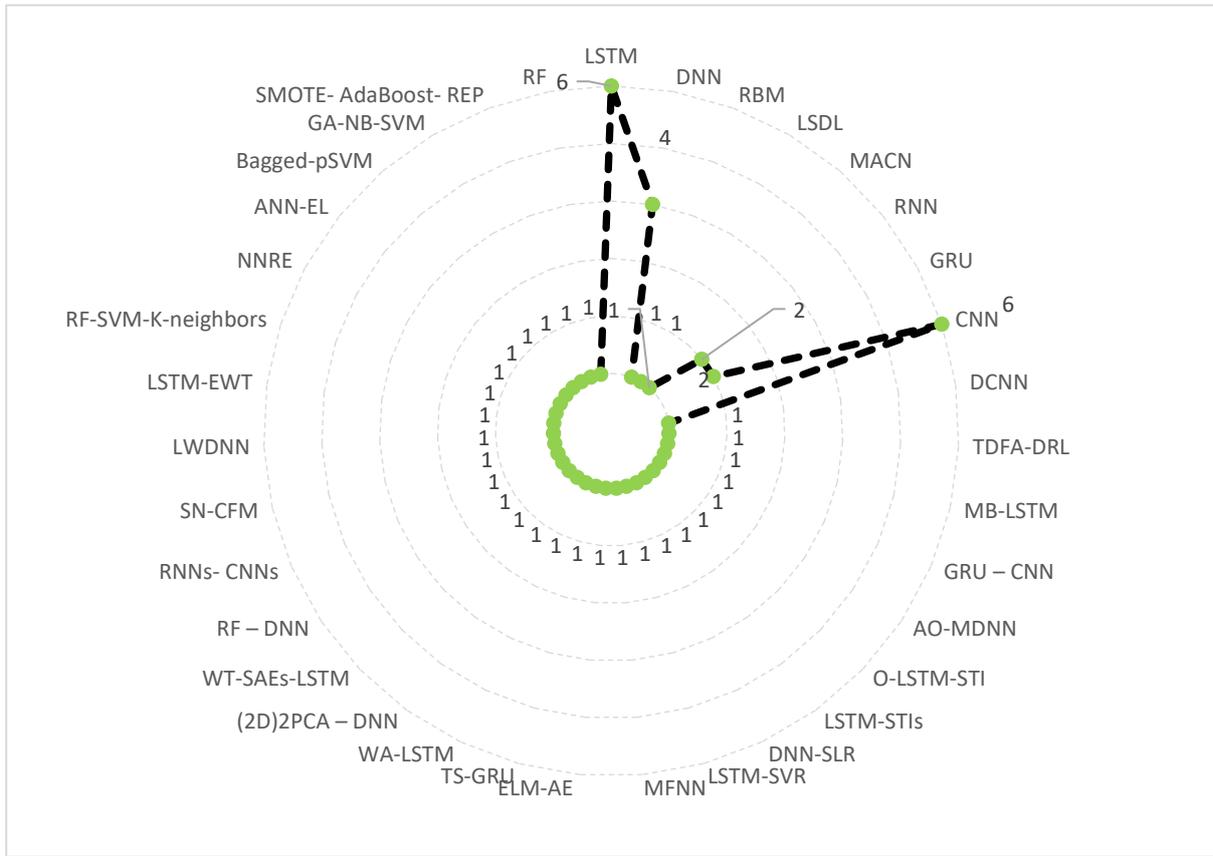

Figure 3. List of all deep learning and hybrid deep learning models applied in Economics related fields

In other words, almost 51% of the models have been hybrid models which represents a trend in the literature to use hybrid models in Economics. This may be because the predictive power of hybrid models is higher than that of single DL models. Therefore, Root Mean Squared Error (RMSE), that is one of the accuracy metrics of machine learning models, of all these models were investigated and shown in Figure 4. Less RMSE, more the accuracy of the model [4]. Figure 4 shows that the RMSE reported for the single DL models have been higher than the hybrid DL models. This finding is consistent with the claims made in the literature (e.g. [3-5]) and it can also explain why the use of hybrid models has become trendy.

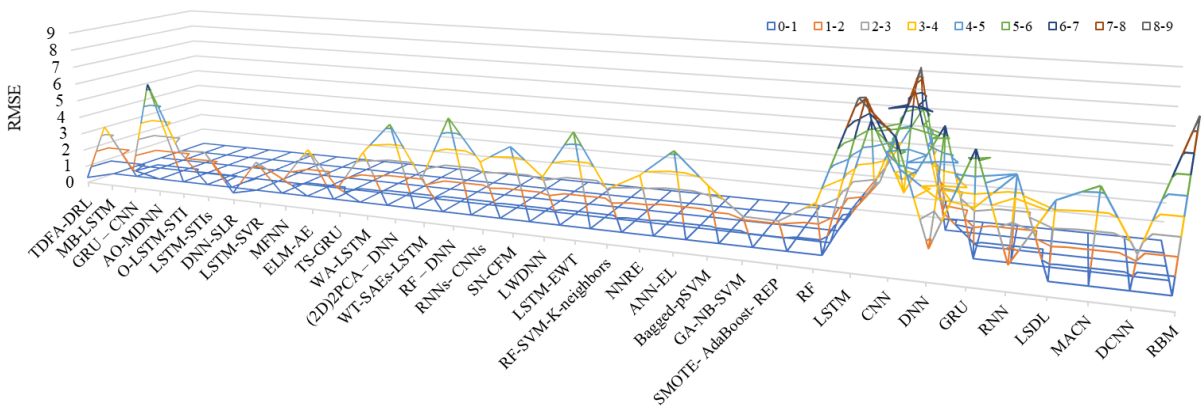

Figure 4. Comparison of RMSE values of hybrid deep learning models and deep learning

## 5. Conclusion

The development of high-precision analysis and prediction algorithms has made the application of data science to various disciplines including disciplines related to economics very attractive. The findings of this study show that, firstly, the advancement of data science algorithms was in three groups of deep learning models, ensemble models, and hybrid models. These three type models were investigated in five different areas, namely 1) stock market, in which stock price prediction was the main goal of most of the papers, 2) marketing, in which the objectives of the papers were mostly to study the customer behavior, 3) corporate bankruptcy, 4) cryptocurrency, which is a new trend in the economic field where the researchers try to predict the price of digital moneys, 5) E-commerce, in which the DL models are applied mainly to increase the performance of e-commerce websites for example by item categorization and product recommendation. This study elaborately provides the algorithms of deep learning models used in the economics, data source, and purpose of data science use in each category. Findings reveal that the use of hybrid models has increased due to the higher prediction accuracy than single deep learning models. LSTM, CNN and DNN models have been respectively the most applied models in the literature to analyze financial time series data and predict stock price. In addition, it is found that more than 51% of the models applied in the database of this research are hybrid models. The results of comparing RMSE values of hybrid deep learning models and deep learning models disclosed that hybrid models show a remarkable low error level, indicating that such models have higher prediction accuracy. Therefore, it is recommended to apply hybrid models to model and optimize objectives in all the four fields of studied in this research. In a nutshell, the contribution of this study is to present the advancement of data science in the three sections of deep learning methods, hybrid deep learning methods, and ensemble machine learning models in Economics related fields. This study provides a ground for the future studies and provide the insights for the practitioners to select the most appropriate model for their specific applications.

**Author Contributions:** Conceptualization, A.M.; investigation, S.N.; writing—original draft preparation, S.N.;

methodology, A.M.; writing—review and editing, A.M. and P.G.; visualization, S.N; validation, A.M.; supervision and controlling the results, P.G.

**Funding:**

**Acknowledgments:**

**Conflicts of Interest:** The authors declare no conflict of interest.